# Stone circles on the harraat of Syrian desert

**Amelia Carolina Sparavigna**
Department of Applied Science and Technology
Politecnico di Torino, C.so Duca degli Abruzzi 24, Torino, Italy

*Recently some stone structures covering the harraat of the Syrian desert had been investigated using satellite imagery (arXiv:1106.4665). Dispersed in an arid landscape, they were probably made during the Neolithic period. Some of these structures are here discussed, which display a possible use as ancient sun observatories. For the analysis, a freely available software is used which can be quite suitable for evaluating the effect of solar radiation on physical features of natural structures too.*

An arid land, characterized by huge lava fields, the "harraat", is covering a large part of the Middle East. It is the Syrian Desert, extending from the northern Arabian Peninsula to the eastern Jordan, southern Syria and western Iraq, and in the past considered as a barrier between Levant and Mesopotamia. This desert has two volcanic regions. One is the Jabal al-Druze, in the As-Suwayda Governorate, consisting of several basaltic volcanoes active from the lower-Pleistocene to the Holocene [1]. The other field is that of the Harrat Ash Shaam. In Arabic, the lava fields are the harraat (sing. harrah; before a name, harrat). Harrat Ash Shaam includes the Es-Safa volcano, in the South of Syria, south-east of Damascus [2].

A survey of this desert with satellite imagery, for instance with the Google Maps, shows the lava fields and the extent of harraat, with different colors of magma extruded in past epochs [3]. We see also that this region is crossed by some huge stone structures, the "desert kites", which were kite-shaped Neolithic stone fences, probably used as animal traps [3-5]. Besides the kites, there are also abundant archaeological evidence of Neolithic communities because thousands of tumuli, stone fences, circular structures are covering large areas [6]. Sometimes we can see some structures superimposed to older ones (see for instance the Fig.1).

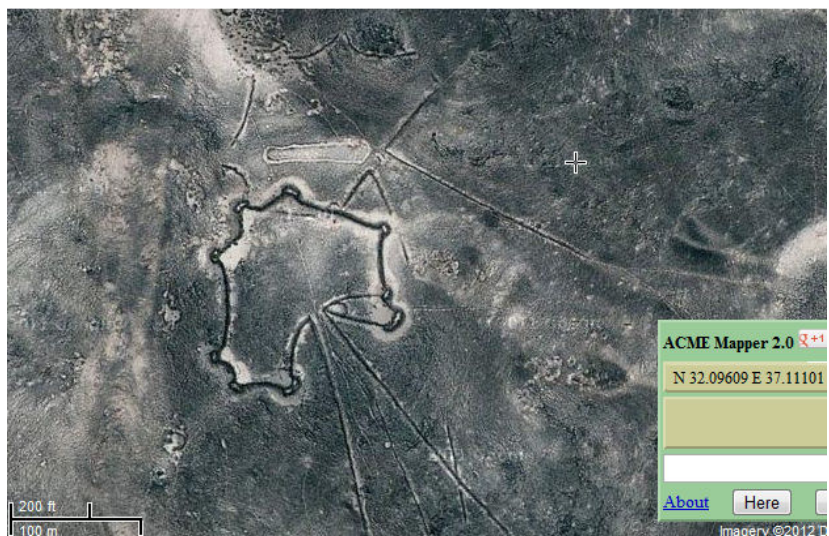
Fig.1 Desert kites.

Several stone circles are visible too, as many Stonehenge sites dispersed in the desert landscape. "Referred to by archaeologists as "wheels," these stone structures have a wide variety of designs, with a common one being a circle with spokes radiating inside." [7]. These structures range from 25 meters to 70 meters across. Proposed as "Stonehenges" of the Middle East [8], and compared with the Nazca lines as possible geoglyphs with astronomical alignments [7], we can investigate whether the hypothesis of being ancient rudimental sun observatories is possible or not.

The prototypes of ancient solar observatories in Europe are the sites of Stonehenge and Goseck and, in Peru, the complex of Chankillo [9-11]. The European sites have generally a circular structure, but Chankillo is a more or less straight line of towers on a low hill. Therefore, a structure connected with the observation of sunrise, sunset and zenith (if any) position of the sun, can have quite different shapes according to its location relative to the Equator, that is, to its latitude.

Recently I have discussed the orientation of ancient temples and roman towns in some paper using the angles of the horizontal and equatorial systems [12-14]. Since the Syrian wheels have clear radial structures, let us use a different approach and compare them with directions of sunset, noon, and sunrise, as given according to their latitude. Some software, developed for solar energy applications, is available for our purpose. Among the many sites providing solar information, we can choose http://www.sollumis.com/, that proposed a model of sunlight direction into a Google map on any day of the year. Figure 1-3 show some of the stone circles of Syrian desert already displayed in Ref.4, analysed according to the solar directions. It seems that there is a good agreement between sunrises and sunsets at equinoxes as given by the model and stone circle features. There is also a site, that shows a certain alignment of a line of stones with the sunrise on the winter solstice as given by the model of sollumis.com (see Fig.4).

At a first glance then, the proposal of Ref. 8 of some "Stonehenges" in the Middle East seems to be reliable. Of course, it is necessary to tell that some hills on the horizon can alter the direction of apparent sunrise and sunset and therefore an on-site evaluation is required. Moreover, the original alignment of stones could had been different from the observed one, that is, it could had been altered during time by the local population, to have an agreement with sunrise and sunset at equinoxes, creating their own rudimental solar observatory. It is therefore difficult to give any conclusion on the age of the stone circle layout by their simple observation from satellites [15].

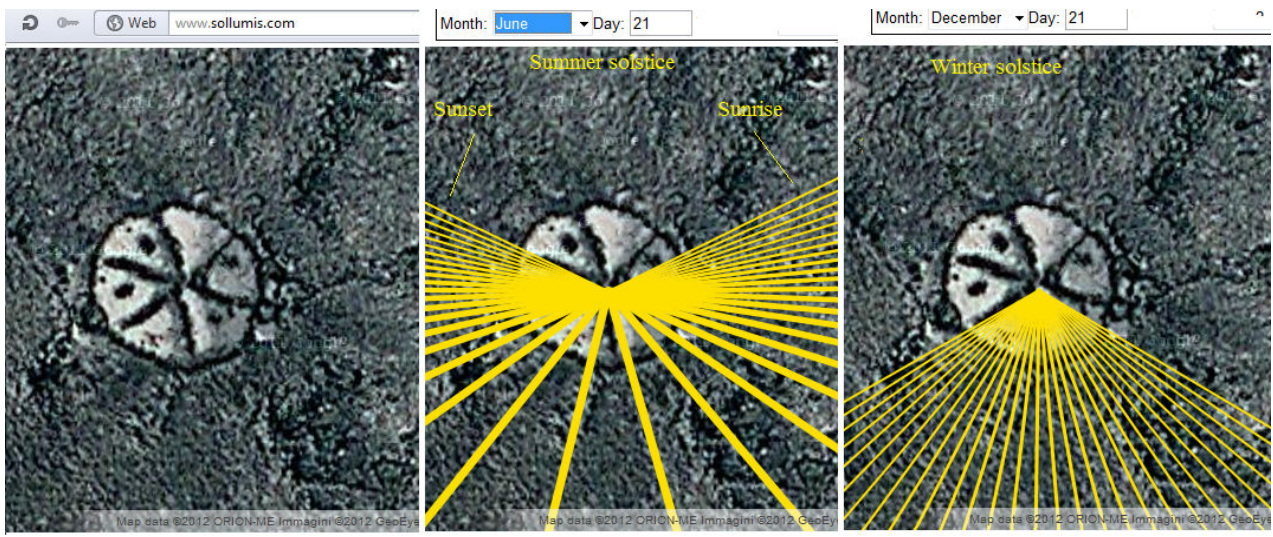

Fig.1 This is one of the stone structure of Syrian Desert. The image shows directions of sun during the day. "The lines on the drawing show the direction and height (altitude) of the sun throughout the day. Thicker and shorter lines mean the sun is higher in the sky. Longer and thinner lines mean the sun is closer to the horizon", according to Sollumis.com http://www.sollumis.com/. On the left, the site as it appears in the Google Maps. In the middle, the direction of the sun on the summer solstice, choosing the center of the circle for observation. We see that, at sunrise, the sun is passing near the dot. At the sunset the direction is that of a line. In the image on the right, we see the direction of the sun on winter solstice. At sunrise, the lines is passing between dots. The sunset has the direction of a radius (images have been obtained from original provided by Google Maps and sollumis.com).

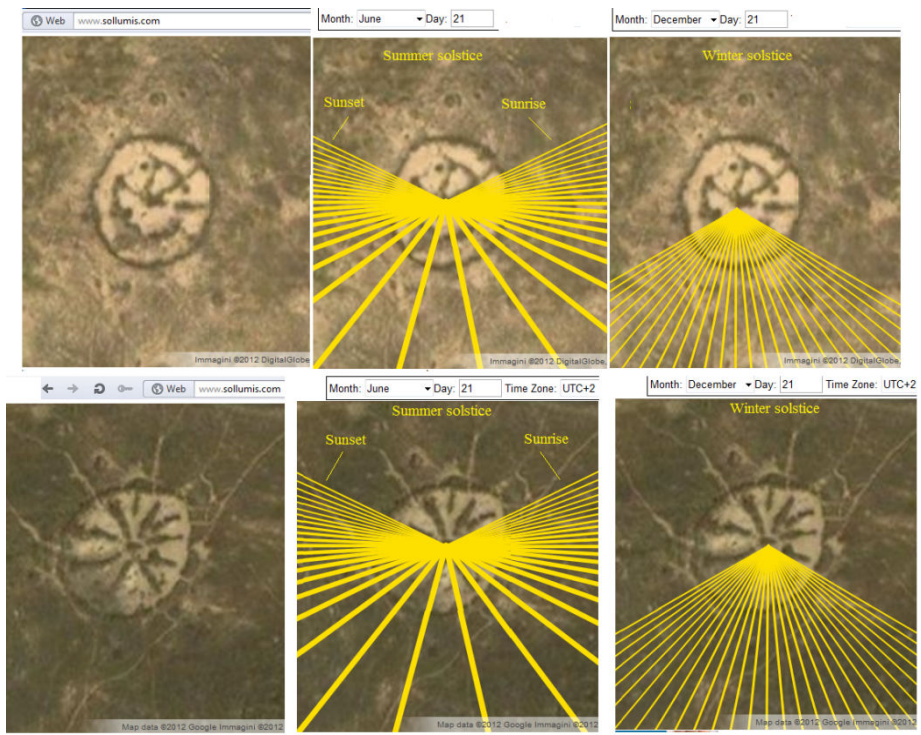

Fig.2 Other two stone circles analysed as in Fig.1, and having, more or less, the same features.

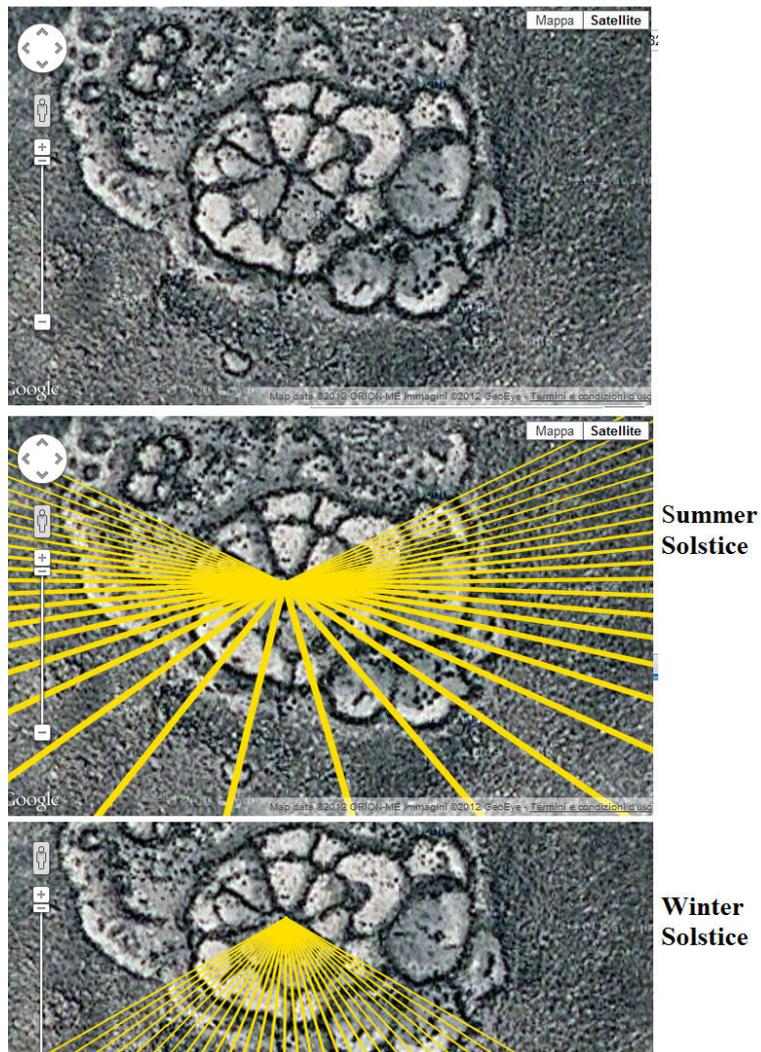

Fig.3 A complex structure in the Syrian desert, with its analysis with sollumis.com

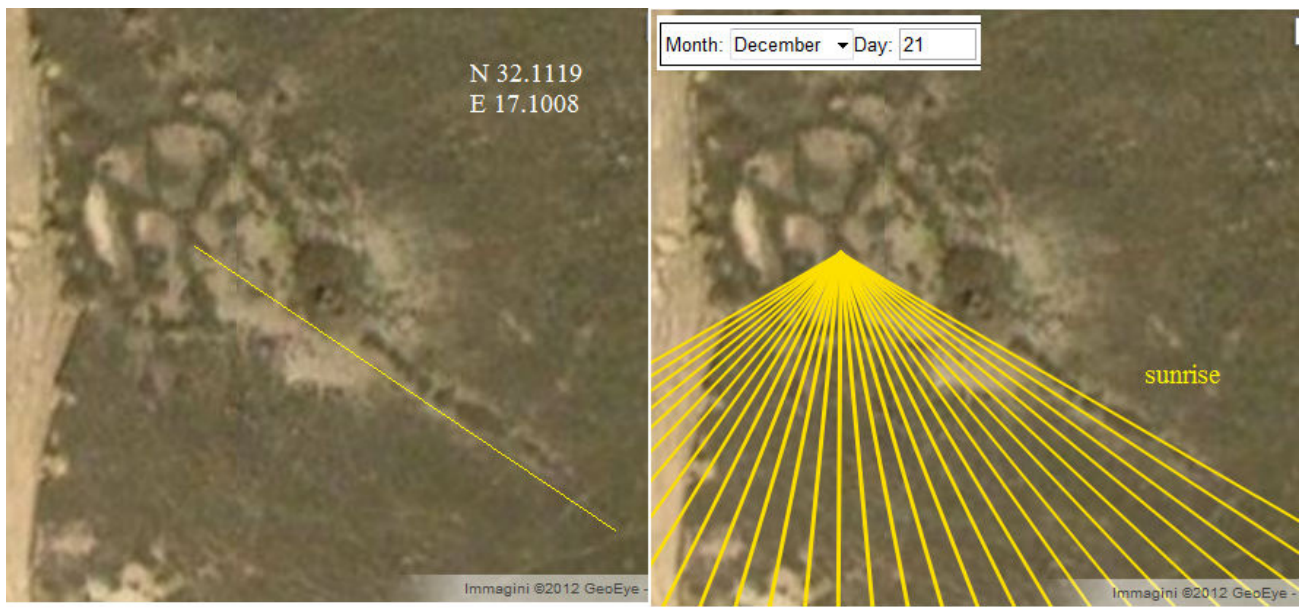

Fig.4 This site has a line of stones which seems aligned with the sunrise of the winter solstice.

Of course, the comparison of structures, natural or human, observed by satellites with models drawing the solar directions, is fundamental for all the solar energy applications. For this reason, it is quite easy to find freely available software on the web. Here we used the software to investigate the orientation of ancient stone circles. But this and other similar software can be applied for modelling the solar radiation on natural structures (lakes, marshes, dunes, etc.) to evaluate their evolution during cycles of years.